# On some consequences of the permutation paradigm for data anonymization: centrality of permutation matrices, universal measures of disclosure risk and information loss, evaluation by dominance[1]

Nicolas Ruiz[2]
OECD

**First version**

## Abstract

Recently, the permutation paradigm has been proposed in data anonymization to describe any micro data masking method as permutation, paving the way for performing meaningful analytical comparisons of methods, something that is difficult currently in statistical disclosure control research. This paper explores some consequences of this paradigm by establishing some class of universal measures of disclosure risk and information loss that can be used for the evaluation and comparison of any method, under any parametrization and independently of the characteristics of the data to be anonymized. These measures lead to the introduction in data anonymization of the concepts of dominance in disclosure risk and information loss, which formalise the fact that different parties involved in micro data transaction can all have different sensitivities to privacy and information.

## 1. Introduction

Data on individual subjects are increasingly collected and exchanged. By their nature, they provide a rich amount of information that can inform statistical and policy analysis in a meaningful way. However, due to the legal obligations surrounding these data, this wealth of information is often not fully exploited in order to protect the confidentiality of respondents. In fact, such requirements shape the dissemination policy of micro data at national and international levels. The issue is how to ensure a sufficient level of data protection to meet releasers' concerns in terms of legal and ethical requirements, while offering to users a reasonable richness of information. Moreover, over the last decade the role of micro data has changed from being the preserve of National Statistical Offices and government departments to being a vital tool for a wide range of analysts trying to understand both social and economic phenomena. As a result, more parties, often very heterogeneous in their privacy and information requirements, are now involved in micro data transactions. This has opened a new range of questions and pressing needs about the privacy/information trade-off and the quest for best practices that can be both useful to users but also respectful of respondents' privacy.

Statistical disclosure control (SDC) research has a rich history in addressing those issues, by providing the analytical apparatus through which the privacy/information trade-off can be assessed and implemented. SDC consists in the set of tools that can enhance the level of confidentiality of any data while preserving to a lesser or greater extent its level of information (see [8] for an authoritative survey). Over the years, it has burgeoned in many directions. In particular, techniques applicable to micro data,

---

[1] The author would like to thank Josep Domingo-Ferrer and Krishnamurty Muralidhar for their comments and suggestions on a preliminary version of this paper.
[2] Contact : nicolas.ruiz@oecd.org. OECD, Rue André Pascal, 75016, Paris, France. Tél.: +33145241433

which are the focus of this paper, offer a wide variety of tools to protect the confidentiality of respondents while maximizing the information content of the data released, for the benefits of society at large.

Streaming from the the large variety of practical cases that can occur in micro data exchange is the diversity of techniques available for data anonymization. Such diversity is undoubtedly useful but has however one major drawback: a lack of agreement and clarity on the appropriate choice of tools in a given context, and as a consequence a lack of general view (or at best an incomplete one) across the relative performances of the techniques available. In fact, the cross-evaluation of current micro data masking methods is a challenging task for at least two reasons. The first is analytical: the evaluation of utility and privacy for each method is metric and data-dependent ([10]). As a result, there is no common language for comparing different mechanisms, all with potentially varying parametrizations applied on the same data set or different data sets. Moreover, there is also a variety of definition for privacy and information loss, and picking some is often related to the context in which they are used and/or can result from an arbitrary choice. The fact that all evaluations can only be practical in nature and context-specific is clearly an issue, not least precluding a sound and simple communication on data anonymization as well as a wider democratization of the field that could allow for more data to be disseminated.

A second reason is related to the variety of parties involved in micro data exchange. Indeed, it is natural to assert that across each party different sensitivities to privacy and information prevail. Some may place greater emphasis on the preservation of privacy, e.g. typically the data releasers, while others are relatively more concerned by the extent to which information is preserved, e.g. typically the researchers. Additionally, these sensitivities can differ also within groups, e.g. one researcher can have a low sensitivity to information loss and consider a release better than no release at all, while another could simply disregard the data above a certain threshold of loss set according to his intended use of the data.

A step toward the resolution of such limitations has been recently proposed ([11] and [3]), by establishing that any micro data masking method can be viewed as functionally equivalent to a permutation of the original data plus eventually a small noise addition. This insight, called the permutation paradigm, unambiguously establishes a common ground upon which any masking method can be gauged. It is independent of the underlying parameters of the masking mechanism and the characteristics of the data. Moreover, it presents the advantage of being meaningful and easy to grasp and implement, as the only necessary and sufficient information for the comparative evaluation of some methods, being under different parametrizations and/or different data sets, is a distribution of permutation distances. Thus, the permutation paradigm is also a tremendous simplifier for data anonymization.

While this paradigm is not considered by its author as a new anonymization method *per se* (a statement that can be reconsidered, see later), it offers the potential to re-interpret all the techniques available through the same lens. It remains however to develop a set of appropriate measures of disclosure risk and information loss based on permutation distances. This is the objective of this paper, which explores some consequences of the permutation paradigm. Notably, it proposes some universal measures of disclosure risk and information loss that can be computed in a simple fashion and used for the evaluation of any anonymization methods, independently of the context under which they operate. The construction of these measures allows introducing in data anonymization the notions of dominance in disclosure risk and information loss, which formalise the fact that different parties involved in micro data release can all have different sensitivity to privacy and information, and can inform about the methods that can reach a consensus among all parties involved. These two notions of dominance can in fact characterize which methods, under any tastes for privacy and information, always perform better than others.

This paper first starts in Section 2 with a brief reminder of the permutation paradigm and one of its first, simple consequence, which establishes permutation matrices as an encompassing tool in data anonymization. From permutation matrices, Section 3 derives a general class of disclosure risk measures and introduces the concept of dominance in disclosure risk. Section 4 then develops a general class of information loss measures as well as the related concept of dominance in information. Section 5 proceeds with possible extensions of the measures introduced in this paper. Finally, conclusions and paths for future research are gathered in Section 6.

## 2. Centrality of permutation matrices in data anonymization
### 2.1 Restatement of the permutation paradigm

The current state of the literature on data anonymization offers a wide variety of techniques suited to different circumstances in terms of data, utility preservation and privacy requirement ([8]). But as outlined above, this diversity in techniques also entails some difficulties in comparing the level of utility and privacy achieved through different methods on different data sets, as all of them are ultimately tied to the analytical framework selected, in particular their parameters which are data-dependent, and the underlying metrics used. This makes the comparison of different mechanisms such as e.g. additive vs. multiplicative perturbations, or the same mechanism applied on different data sets, an awkward task. However, a recent contribution in the literature (see [11] and its subsequent development in [3]) proposed a general functional equivalence to describe any data masking method. From the observation that any anonymized data set can be viewed as a permutation of the original data plus a non-rank perturbative noise addition, the authors established that all masking methods can be thought of in term of a single ingredient, i.e. permutation. The so-called permutation paradigm has clearly far reaching conceptual and practical consequences, in the sense that it provides a single and easily understandable reading key, independent of the model parameters, the risk measures or the specific characteristics of the data, to interpret the utility/protection outcome of an anonymization procedure.

To illustrate the permutation paradigm, we introduce a simple running example which consists (without loss of generality) of five records and three attributes $X=(X_1, X_2, X_3)$ generated by sampling $N(10,10^2)$, $N(100,40^2)$ and $N(1000,2000^2)$ distributions, respectively. Noise is then added to obtain $Y=(Y_1, Y_2, Y_3)$, the three masked version of the attributes, from $N(0,5^2)$, $N(0,20^2)$ and $N(0,1000^2)$ distributions, respectively. One can see that the masking procedure generates a permutation of the records of the original data (Table 1).

**Table 1. An illustration of the permutation paradigm**

| Original dataset X | | | Masked dataset Y | | |
|---|---|---|---|---|---|
| $X_1$ | $X_2$ | $X_3$ | $Y_1$ | $Y_2$ | $Y_3$ |
| 13 | 135 | 3707 | 8 | 160 | 3248 |
| 20 | 52 | 826 | 20 | 57 | 822 |
| 2 | 123 | -1317 | -1 | 122 | 248 |
| 15 | 165 | 2419 | 18 | 135 | 597 |
| 29 | 160 | -1008 | 29 | 164 | -1927 |
| Rank of the original attribute | | | Rank of the masked attribute | | |
| $X_{1R}$ | $X_{2R}$ | $X_{3R}$ | $Y_{1R}$ | $Y_{2R}$ | $Y_{3R}$ |
| 4 | 3 | 1 | 4 | 2 | 1 |
| 2 | 5 | 3 | 2 | 5 | 2 |
| 5 | 4 | 5 | 5 | 4 | 4 |
| 3 | 1 | 2 | 3 | 3 | 3 |
| 1 | 2 | 4 | 1 | 1 | 5 |

Now, as long as the attributes' values of a dataset can be ranked, which is obvious in the case of numerical and categorical ordinal attributes, but also feasible in the case of nominal ones ([5]), it is always possible to derive a dataset Z that contains the attributes $X_1$, $X_2$ and $X_3$, but ordered according to the ranks of $Y_1$, $Y_2$ and $Y_3$, respectively, i.e. in Table 1 re-ordering $(X_1, X_2, X_3)$ according to $(Y_{1R}, Y_{2R}, Y_{3R})$. This can be done following the post-masking reverse procedure outlined in [3]. Finally, the masked data Y can be fully reconstituted by adding small noises $(E_1, E_2, E_3)$ (small in the sense that they cannot re-rank Z while they can still be large in absolute values) to each observation in each attribute (Table 2).

Table 2. Equivalence in anonymisation: postmasking reverse mapping plus noise addition

| Original dataset X | | | Reverse mapped dataset Z | | |
|---|---|---|---|---|---|
| $X_1$ | $X_2$ | $X_3$ | $Z_1$ | $Z_2$ | $Z_3$ |
| 13 | 135 | 3707 | 13 | 160 | 3707 |
| 20 | 52 | 826 | 20 | 52 | 2419 |
| 2 | 123 | -1317 | 2 | 123 | -1008 |
| 15 | 165 | 2419 | 15 | 135 | 826 |
| 29 | 160 | -1008 | 29 | 165 | -1317 |

| Noise E | | | Masked dataset Y(=Z+E) | | |
|---|---|---|---|---|---|
| $E_1$ | $E_2$ | $E_3$ | $Y_1$ | $Y_2$ | $Y_3$ |
| -5 | 0 | -459 | 8 | 160 | 3248 |
| 0 | 5 | -1597 | 20 | 57 | 822 |
| -3 | 0 | 1256 | -1 | 122 | 248 |
| 2 | 0 | -229 | 18 | 135 | 597 |
| 0 | -1 | -610 | 29 | 164 | -1927 |

By construction, Z has the same marginal distribution as X, which is an appealing property. Moreover, under a maximum-knowledge intruder model of disclosure risk evaluation, the small noise addition turns out to be irrelevant ([3]): re-identification *via* record linkage can only come from permutation, as by construction noise addition cannot alter ranks. Reverse mapping thus establishes permutation as the overarching principle of data anonymization, allowing the functioning of any method to be viewed as the outcome of a permutation of the original data, independently of how the method operates. This functional equivalence leads to the following proposition:

**Proposition 1**: *For a dataset[3] $X_{(n,p)}$ with n records and p attributes ($X_1,..,X_p$), its anonymized version $Y_{(n,p)}$ can always be written, regardless of the anonymization methods used, as:*
$$Y_{(n,p)} = (P_1 X_1, \ldots, P_p X_p)_{(n,p)} + E_{(n,p)}$$
*where $P_1,..,P_p$ is a set of p permutation matrices and $E_{(n,p)}$ is a matrix of small noises.*

In the example of Table 2 one can indeed easily verify that:

$$P_1 = \begin{pmatrix} 1 & 0 & 0 & 0 & 0 \\ 0 & 1 & 0 & 0 & 0 \\ 0 & 0 & 1 & 0 & 0 \\ 0 & 0 & 0 & 1 & 0 \\ 0 & 0 & 0 & 0 & 1 \end{pmatrix} \quad P_2 = \begin{pmatrix} 0 & 0 & 0 & 0 & 1 \\ 0 & 1 & 0 & 0 & 0 \\ 0 & 0 & 1 & 0 & 0 \\ 1 & 0 & 0 & 0 & 0 \\ 0 & 0 & 0 & 1 & 0 \end{pmatrix} \quad P_3 = \begin{pmatrix} 1 & 0 & 0 & 0 & 0 \\ 0 & 0 & 0 & 1 & 0 \\ 0 & 0 & 0 & 0 & 1 \\ 0 & 1 & 0 & 0 & 0 \\ 0 & 0 & 1 & 0 & 0 \end{pmatrix}$$

and $E_{(n,p)}$ is given by the lower-left matrix of table 2.

*Proposition 1* is simply a restatement of the permutation paradigm. It has however several implications. The first is that it characterises permutation matrix as an encompassing tool for data anonymization: the analytical framework of anonymization mechanisms can in fact be viewed as functionally equivalent to a set of permutation matrices. Clearly, this formalizes the common basis of comparison for different mechanisms that the permutation paradigm originally proposed. Whatever the differences in the natures of the methods to be compared and the distributional features of the original data, the methods can fundamentally always be viewed as the application of different permutation matrices to the original data. And as a standard tool in linear algebra, permutation matrices are meaningful, readable and practical in comparison to the sometimes quite complex analytical apparatus of some masking methods.

---

[3] In the remainder of this paper, the subscript in parenthesis describes the number of rows and columns for a matrix. Here for example $X_{(n,p)}$ is a matrix with n rows and p columns.

Second, it is clear that $P_1,..,P_p$ are independent of the data characteristics, as each permutation matrix can be dealt with in isolation of the data set to be anonymized. So while *Proposition 1* doesn't describe a new anonymization method, but instead a way of seeing *any* anonymization method, nothing precludes, conceptually or practically, to think about data anonymization only in term of permutation. Approaching data anonymization in this way could offer several advantages. Given the fact that this can be done independently of the data, one could consider for example data releasers to be equipped with an arsenal of permutation matrices, to be applied to different data sets and configurations of data utility/privacy streaming from the demands that are addressed to them, or to the different stringency of the rule of law for data dissemination that they face. Also, ex-post evaluation of disclosure risk (in a utility-first approach to anonymization) or of data utility (in a privacy-first approach to anonymization), which can lead to several re-runs of methods to reach the appropriate settings, can potentially all be carried out ex-ante in the permutation paradigm. This has the potential to provide a more efficient and less costly approach to anonymization for data releasers.

### *2.2 A new roadmap for data anonymization*

The broad conceptual implication of the permutation paradigm can also potentially pave the way for changes in the way data anonymization is practiced. Now, any procedure used can be systematically translated into the permutation paradigm to report the outcomes in terms of permutation. Indeed, several paths for future research were proposed by the original authors of the paradigm ([11], [3]). Among them, this paper addresses particularly the formal characterisation of permutation distances and the derivation of appropriate disclosure risk and information loss measures. We argue that the proposals put forward in the remainder of this paper can fit in a new general scheme for data anonymization, where current methods (and their different parametrizations) can all be judged through the same lens (Figure 1).

**Figure 1. Data anonymisation roadmaps**

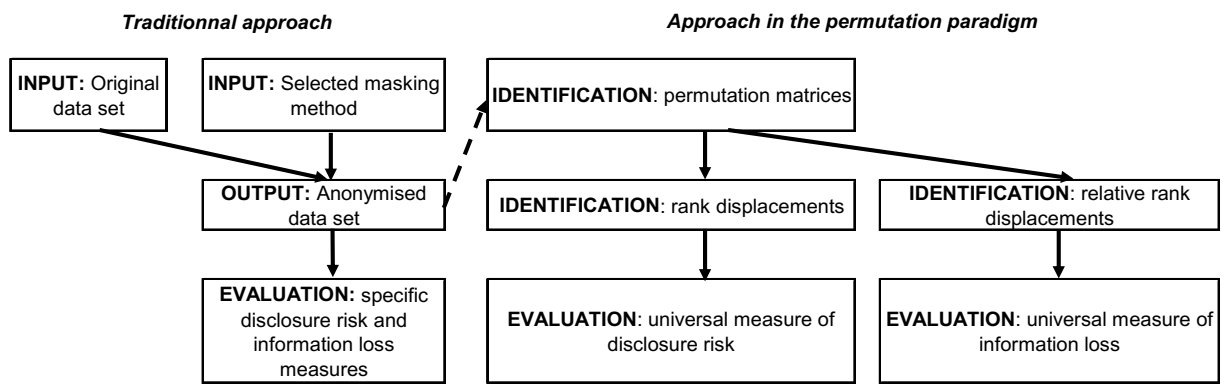

The current practice is schematised on the left of Figure 1, where a data set is anonymized according to a given method in order to deliver the masked version of the data set. This masked version is then evaluated in terms of disclosure risk and information loss according to currently existing metrics, which are generally specific to the environment. In the permutation paradigm, on the right of Figure 1[4], permutation matrices are identified from the anonymized data set, from which can be retrieved, as will be developed below, vectors of rank displacements and relative rank displacements, which can in turn be used to compute universal measures of disclosure risk and information loss. We qualify these measures as universal for two reasons. The first is that they allow performing comparison across different methods and data sets. They are not specific to the environment. The second is that they can incorporate different judgments on disclosure risk and information loss. In that sense, they are able to account for the different preferences that can co-exist in a data exchange process.

---

[4] Note that for the same reasons outlined above, permutation matrices can potentially be considered as a direct input for anonymization, replacing the choice of method.

### 3. A class of universal measures of disclosure risk based on permutation distances

As we saw, permutation matrices offer a single metric from which data utility and privacy can be assessed. From such matrices we start by formalizing a preliminary, simple measure of disclosure risk:

**Proposition 2**: *For any attribute j=1,...,p of $Y_{(n,p)}$, a qualitative measure of disclosure risk in the permutation paradigm is given by:*

$$T_j = 1 - \frac{trace(P_j)}{n}$$

*with $T_j=1$ when all records have been permuted at least one time (low disclosure risk), and 0 when no permutation occurred (high disclosure risk; in that case $P_j$ is the identity matrix).*

The trace of a permutation matrix is the number of fixed points in the permutation. Thus, $T_j$ measures the extent of permutation of attribute j in a qualitative way, in the sense that it informs about permutations as a proportion of the total number of records, but doesn't convey any information on the magnitudes of the permutations. It allows nonetheless in a simple fashion to do a first screening of protection against disclosure risk. In the example of Table 2, $T_1=0$, $T_2=0.6$ and $T_3=0.8$; of the three attributes, the first has no protection against disclosure risk while the third has the highest protection, according to $T_j$.

While the overall amount of permutation performed by an anonymization method matters, permutation distances are crucial to assess protection against disclosure risk; even if all records have been permuted, the closest they have been the higher will be disclosure risk. For a given attribute j and its associated permutation matrix $P_j$, permutation distances can be retrieved by the computation of a vector of rank displacement $r_j$, i.e. a vector measuring for each record the amount of rank shifting that a permutation matrix contains. Note that to avoid some unnecessary technical difficulties, in what follows zero values in $r_j$ will be assigned, without loss of generality, a infinitesimally small value $\varepsilon>0$.

In order to build $r_j$, one can count, columns by columns of $P_j$, how many times the 1s have been moved, using the identity matrix as a starting point (which is a particular case of a permutation matrix with no permutation applied), then assigning a negative (resp. positive) sign if the 1 has been moved up (resp. down). For the matrices $P_1$, $P_2$ and $P_3$, one gets:

$$r_1 = \begin{pmatrix} \varepsilon \\ \varepsilon \\ \varepsilon \\ \varepsilon \\ \varepsilon \end{pmatrix} \quad r_2 = \begin{pmatrix} 3 \\ \varepsilon \\ \varepsilon \\ 1 \\ -4 \end{pmatrix} \quad r_3 = \begin{pmatrix} \varepsilon \\ 2 \\ 2 \\ -2 \\ -2 \end{pmatrix}$$

Now, $r_j$ has to be evaluated in some way for assessing disclosure risk based on permutation distances. A natural choice is to gauge $r_j$ by assigning a magnitude, taking its Euclidean norm and adopting the rule that the higher the norm, the lower the disclosure risk (as the larger will be the permutation distances contained in $r_j$). But other cases are possible. In general, any L(p)-norm is acceptable: for example, for $r_1$, $r_2$ and $r_3$, the ∞-norm (or Chebyshev distance) would give $\varepsilon$, 4 and 2, respectively. This variety of choice to evaluate vectors generally depends on the problem at hand, as one will select a L(p)-norm adapted to the meaning of the object that is meant to be quantified. In the case of a vector of permutation distances, it is not clear why a Euclidean length would be more suitable and meaningful than a Chebyshev length, or why all the norms in-between can or cannot be considered. Thus, there can be a fundamental arbitrariness in this choice. However, we argue that in the permutation paradigm, such choice can be given an intuitive interpretation in term of disclosure risk.

To illustrate further this arbitrariness, consider the following example: if in $r_3$ the third record is now permuted one rank more and the second one rank less, $r_3$ will be viewed as identical to $r_2$ according to the ∞-norm. It is however not totally clear if the situation has really improved in term of disclosure risk for the third attribute. On the contrary, it can be reasonably considered that the new situation is more problematic as having now a record permuted only one time increases the disclosure risk in a way that may not be offset by the additional permutation of an already sufficiently permuted record. In fact, being able to evaluate if the situation has improved necessitates a notion of aversion to disclosure risk, which, to the best of the author knowledge, is not present or formalized in the literature on SDC. The permutation paradigm allows in a simple way to introduce this notion:

***Definition 1***: *In the permutation paradigm, aversion to disclosure risk is the preference toward less permuted records for the evaluation of this risk.*

Aversion to disclosure risk accounts for the fact that different data releasers or subjects can all have potentially different appreciations of disclosure risk (alternatively, this can also be viewed as different levels of privacy awareness). Some releasers may consider that achieving a certain average level of permutation is sufficient, while from a contributing subject's point of view, or from the point of view of other data releasers (say for example when several ones are involved in the release of a data set), this could be judged as not enough. Because the permutation paradigm reduces the relevant information needed for the evaluation of any method to permutation, aversion to disclosure risk can be modelled by assuming that different permutation distances have different weights. On the one hand, a strongly averse data releaser/subject may put relatively more weight on the lowest permutation distances achieved; on the other hand, a weakly averse releaser/subject may consider different permutation distances the same way and focuses only on the average amount of permutations.

In fact, already existing measures of disclosure risk entail generally some implicit assumptions regarding how the risk is assessed. This can be illustrated by considering the formula for rank order correlation coefficient, previously used in the permutation paradigm for the assessment of disclosure risk ([11], [3]), which for a non-masked attribute $X_j$ and its reverse mapped version $Z_j$ can be written as (where $d_i$ is the difference between the ranks of each record):

$$\rho_{X_j,Z_j} = 1 - \frac{6\sum_{i=1}^n d_i^2}{n(n^2-1)}$$

It is apparent that the rank order correlation coefficient implies specific preferences on the permutation distances, as the square of the ranks' differences magnifies the impact of large permutations compared to small ones. One could even argue that the rank order correlation coefficient is not an appropriate measure, as for the assessment of disclosure risk it is small, not large, permutation distances that matter. For example, according to $\rho_{X_j,Z_j}$ an anonymization method permuting only one record 10 times will be judged as having reduced disclosure risk more than another method permuting 3 records 5 times. Once again, it is hard to rank the two situations in terms of disclosure risk. To overcome this, the following proposition establishes a measure of disclosure risk sensitive to different aversions, with an adjustable degree of focus on small permutation distances:

***Proposition 3***: *For any attribute j=1,…,p of $Y_{(n,p)}$, a quantitative measure of disclosure risk in the permutation paradigm is given by:*

$$D_j(\alpha) = \left[\frac{1}{n}\sum_{i=1}^n abs(r_{j(i)})^\alpha\right]^{1/\alpha} \quad for\ \alpha \leq 1\ and\ \alpha \neq 0$$

$$and\ D_j(\alpha) = \prod_{i=1}^n abs(r_{j(i)})^{1/n} \quad for\ \alpha = 0$$

*where $r_{j(i)}$ denotes the elements of $r_j$ and α the parameter of aversion to disclosure risk.*

$D_j(\alpha)$ makes use of a power mean[5] (see [7] for a discussion of its various properties) for the aggregation of the components of $r_j$, with the parameter $\alpha$ substantiating the notion of aversion to disclosure risk. The arithmetic mean becomes a special case ($\alpha = 1$) of $D_j(\alpha)$, which forms a natural starting point by computing the average level of permutation distances. In that case, all distances are given the same weight and there is a one-to-one substitution between them, e.g. two records permuted two ranks are equivalent to one record permuted four ranks. From this benchmark, the more α decreases, the more weight is given to the smallest permutation distances[6]. In fact, the more α approaches -∞, the more $D_j(\alpha)$ converges towards the smallest permutation distance in $r_j$[7]. As a result, for a given $r_j$ and $\alpha' < \alpha$, we have $D_j(\alpha') \leq D_j(\alpha)$: the lower is α, the stronger is the aversion to disclosure risk. Table 3 below illustrates some computations of $D_j(\alpha)$ on the running example:

**Table 3. Example of quantitative measures of disclosure risk in the permutation paradigm**

| Aversion to disclosure risk | $r_1$ | $r_2$ | $r_3$ |
|---|---|---|---|
| α=1 | 0 | 1.60 | 1.60 |
| α=0.5 | 0 | 0.97 | 1.06 |
| α=-4 | 0 | 0 | 0 |

Note: ε is assumed equal to 1E-08

For attribute 1, the anonymization method used doesn't protect against disclosure risk, as $D_1(\alpha)$ is equal to zero for any α. For attribute 2 and 3, the average level of permutation distances is the same, i.e. $D_2(\alpha) = D_3(\alpha) = 1.6$. However, different levels of aversion to disclosure risk lead to different diagnoses: for a mild aversion (α = 0.5) the third attribute is judged to be better protected than the second while for a strong aversion (α = −4) the two attributes re-become equivalent. Note that as a general case of average, $D_j(\alpha)$ is independent of the number of records, which eases the comparison across different data sets. Moreover, for an attribute of n records, the maximum permutation distance for a record is abs(n-1). Thus, re-scaling $D_j(\alpha)$ by 1/n-1 will produce a measure of risk that ranges between 0 and 1, which is an appealing property for performing comparisons and quantifying the utility/privacy trade-off ([8]).

The reader might be tempted to think that the notion of aversion to disclosure risk adds an unnecessary layer of complexity to the evaluation of this risk. We argue however that it provides a better grasp with the reality of micro data exchange. In the current state of the literature, it is not a notion that can be made analytically tractable in a straightforward way for most methods (or as we saw is embodied implicitly rather than explicitly). But in the permutation paradigm, permutation distances are the only meaningful quantities under scrutiny, which makes natural the fact that these distances can be judged by different individuals differently. Given the number of parties implied in data dissemination, e.g. several data releasers and respondents, it is very unlikely that all of them will have the same judgment. The $D_j(\alpha)$ measures are a way to incorporate this diversity. In practice, by computing the measure for several α, a data releaser can for example communicate on the prevention against disclosure risk through different points of view. This circumvents the issue involved in the empirical assessment of disclosure risk (see [10]), where a score based on different measures of disclosure risk is computed using an ad-hoc weighting scheme. Under such approach, weights can drive the overall assessment that is made. But using the current proposal, a single measure can be computed on a continuum of weights which all carry an interpretation in term of disclosure risk.

The measure $D_j(\alpha)$ can also be used to characterize in an unambiguous way which data anonymization methods perform better than others through the concept of disclosure risk dominance that we introduce below. The concept of dominance comes originally from the notion of stochastic

---

[5] In linear algebra power mean is also the formula for the computation of p-norms ([1]).

[6] $D_j(0)$ is the geometric mean and $D_j(-1)$ the harmonic mean.

[7] The limit case $D_j(-\infty)$ is strictly equal to the shortest permutation distance in $r_j$.

dominance (see [9]), which is widely used in economics (in particular for the study of risk and inequality). It can however be applied to any distribution, which is done here for the distribution of permutation distances. To the best of the author's knowledge this is the first time it is considered in the context of data anonymization:

**Definition 2**: *For an attribute j, an anonymization method A is said to dominate (i.e. unanimously performs better than) another method B for the protection against disclosure risk if it holds that $D_j(\alpha)' \leq D_j(\alpha) \; \forall \; \alpha \leq 1$ (where $D_j(\alpha)$ (resp. $D_j(\alpha)'$) are the measures of Proposition 3 computed from A (resp. B)).*

Disclosure risk dominance characterizes anonymization methods that will consistently ensure greater levels of permutation distances (and thus levels of protection against disclosure risk) from the mean to the bottom of their distribution. In practice, that means that whatever the aversion any agents being involved in data dissemination can have, a dominant method will ensure unanimity regarding its performance against disclosure risk; in the running example, the third attribute can be judged as having a greater level of protection than the second for any level of aversion (Table 3).

Obviously, dominance may not always be reached in practice. For example, a method A can happen to dominate B over $-4 \leq \alpha \leq 1$ but being dominated by B over $-\infty \leq \alpha < -4$. In that case, that means that the use of A is advisable for small up to medium disclosure risk aversion, while for strong aversion B is more advisable. As a result, on can learn on the relative performance of methods by investigating where dominance holds but also eventually where it ceases to hold.

One final remark is in order on $D_j(\alpha)$. The domain of variation of the disclosure risk aversion parameter has been set to range from one and below, which doesn't define a L(p)-norm strictly speaking. In fact, it would be $D_j(\alpha)$ with $\alpha > 1$ that would rigorously define a L(p)-norm, up to a factor $\sqrt[\alpha]{n}$ (see [1]), leading to a standard notion of distance for the vector $r_j$. However, we argue that in the context of data anonymization, the interpretation of the parameter α is not suited in that case. With $\alpha > 1$, the more α increases, the more weight is given to the largest permutation distances (and the more α approaches $+\infty$, the more $D_j(\alpha)$ converges towards the largest permutation distance in $r_j$, i.e. a Chebyshev distance is computed). That would mean that large permutations make up for the bulk of protection against disclosure risk, but it is small permutations that can lead to greater disclosure risk. As a result, $D_j(\alpha)$ makes use of the aggregation structure of a p-norm but doesn't define one strictly. This has no incidence on the validity and interpretation of the measure.

In this section, the measure $D_j(\alpha)$ and the concept of dominance have been introduced with the aim of offering a more granular view of disclosure risk, with an easy-to-grasp notion of disclosure risk aversion. Given that in the permutation paradigm all the necessary information is reduced to permutation distances as conveyed by the underlying permutation matrices, they provide a common and understandable language for performing meaningful comparisons of anonymization methods, independently of their analytical environment or the distributional features of the data. The class of $D_j(\alpha)$ measures formalizes the tool for such comparisons and is very general in its scope, in the sense that it allows to incorporate different judgments about disclosure risk and to characterise methods that can be viewed as unanimously better than others. From these proposals more measures can be elaborated. For example, one could think in combining $T_j$ and $D_j(1)$ to get:

$$TD_j = T_j * D_j(1)$$

$TD_j$ measures the average permutation distances for attribute j, but discounted by the proportion of non-permuted records. $TD_j$ would then force data releasers to focus both on distances but also on the number of records permuted, which is an alternative way to gauge disclosure risk in the permutation paradigm.

The next step of this paper is to establish a general class of information loss measures. A key feature of the permutation paradigm is that it preserves exactly the marginal distributions of the data (as Z is simply a permutation of X). Thus, information loss can only come from the alteration of the dependency among attributes. This necessitates a view on how multivariate anonymization methods operate in the permutation paradigm.

## 4. A class of universal measures of information loss based on relative permutation distances

We start by adapting *Proposition 1* to outline how a multivariate distribution can be preserved in the permutation paradigm:

***Proposition 4****: For a dataset $X_{(n,p)}$ with n records and p attributes $(X_1,..., X_j, ..., X_{j'}, ...X_p)$, its anonymized version $Y_{(n,p)}$ preserving exactly the joint distribution of $X_j$ and $X_{j'}$ can always be written, regardless of the anonymization methods used, as:*

$$Y_{(n,p)} = (P_1 X_1, \ldots, P_k X_j, \ldots, P_k X_{j'}, \ldots, P_p X_p)_{(n,p)} + E_{(n,p)}$$

*where $P_1,..., P_k,...,P_p$ is a set of p-1 distinct permutation matrices and $E_{(n,p)}$ is a matrix of small noises.*

In the permutation paradigm, the exact preservation of multivariate distributions (here bivariate distributions) can be achieved by applying to some block of attributes the same permutation matrix. In *Proposition 4* the block of X formed by the two attributes j and j' is permuted with the same matrix $P_k$[8] and the joint distribution of the block is exactly preserved. This proposition shows that any multivariate anonymization method can be viewed as a block permutation of attributes. It is a simpler view in comparison to the current multivariate anonymization methods available in the literature, which can be analytically complex (see [8]). Of course, the exact preservation of a multivariate distribution may impinge on the level of privacy achieved by the anonymized data. Additionally, it has been previously established empirically that obtaining a safe anonymized data set that can resist to an attack *via* record linkage necessitates an amount of masking (or equivalently of permutations) proportional to the dependency between the attributes of the original data set ([4]). Expressed in the permutation paradigm, that means that the more dissimilar have to be the permutation matrices.

In practice then, the question turns out to be more about the extent of preservation of multivariate distributions and an inescapable trade-off: the less preservation there is, the more the anonymized data set will be judged as safe. For a dataset with a strong dependence between its attributes, the trade-off may be particularly arduous. But for a dataset with weak attributes dependence it is also a non-trivial issue, as it can be conceived (while very less likely to occur in practice) that an anonymization method can create an artificial dependence between the attributes, which in a way is also a loss of information. For example, it is possible that two completely independent attributes in the original data happen to be, through a peculiar permutation, both ranked in increasing order of magnitudes in the anonymized version, fooling the data user on the real strength of the relationship.

To assess information loss, a first avenue is to compare the rank order correlations between attributes j and j' in the anonymized data and the original data set ([11]). The most likely case is that the former will be lower than the latter, indicating an alteration of the attributes' relationship and thus a loss of information by a weakening of the dependence (but in less likely cases the reverse can also happen). For such comparison, the original level of rank order correlation provides the starting point from which information loss is assessed. As a result, it will differ according to each couple of attributes considered, which is not really convenient. Also, and for the same reason outlined above, an implicit and specific weighting structure is given to large ranks differences when using rank order correlation. Again, different data users can have different views about distances when assessing information loss. As for disclosure risk, this can be formalized through the concept of aversion to information loss (or stated otherwise, of information awareness):

***Definition 3****: For two attributes j and j' in the permutation paradigm, aversion to information loss is the preference toward large relative permutation distances for the evaluation of this loss.*

Thus, a more general approach is to consider the degree of similarity between the permutations that took place for the two attributes and allowing different weights for different *relative* distances. To do

---

[8] It can be noted that substituting $P_k$ by the identity matrix leads exactly to the same result: the other attributes in X are in that case permuted around j and j', who then remain fixed.

so, it can be observed that a vector $\Delta(r_k)$ of differences between the vectors $r_j$ and $r_{j'}$ is a vector of dissimilarity between the anonymization procedures that have been applied to the couple of attributes k=(j, j') (with j≠ j'). When each of the components of $\Delta(r_k)$ are equal to zero, this depicts the case of *Proposition 4* with j and j' having been permuted the same way; the permutation matrices applied to them are identical, despite the fact that the anonymization methods used can be different in practice. There is no loss of information as the joint distribution of j and j' is preserved. But when $\Delta(r_k)$ has some non-zero elements information has been modified. This leads to the following proposition:

**Proposition 5**: *For two attribute j and j' of $Y_{(n,p)}$, a quantitative measure of information loss in the permutation paradigm is given by:*

$$I_k(\theta) = \left[\frac{1}{n}\sum_{i=1}^{n} abs(\Delta r_{k(i)})^\theta\right]^{1/\theta} \quad for\ \theta \geq 1$$

where $\Delta r_{k(i)}$ denotes the elements of $\Delta(r_k)$ and $\theta$ the parameter of aversion to information loss.

The measure $I_k(\theta)$ bears strong analytical similarities with $D_j(\alpha)$, but while the latter is concerned about average or small permutation distances across records for a given attribute, the former considers average or large *relative* permutation distances between two attributes across records. On the running example, we have $\Delta(r_{(2,3)})=\begin{pmatrix}3\\-2\\-2\\3\\-2\end{pmatrix}$, which thus gives $I_{(2,3)}(1) = 2$, $I_{(2,3)}(4) = 2,49$ and $I_{(2,3)}(+\infty) = 3$. Note that this measure delivers a diagnosis independently of the direction of the alteration of dependence between attributes, i.e. if dependence has been weakened or strengthened as a result of anonymization. $I_k(\theta) = 0$ means no information loss while for a given $\theta$, the larger is $I_k(\theta)$, the more the relationship between attributes has been altered (and thus the more information has been lost in the process). It thus provides a general measure of information loss than can be applied to any anonymization methods. Note that $I_k(\theta)$ is a power mean but also denotes strictly a L(p)-norm of the vector $\Delta(r_k)$ up to the factor $\sqrt[\theta]{n}$. This factor allows performing comparison independently of the size of the data set. Moreover, for two attributes with n records each, the maximum relative permutation distance for a record is n-1. Thus, re-scaling $I_k(\theta)$ by 1/n-1 will produce a measure of information loss that ranges between 0 and 1, which is convenient for comparison with $D_j(\alpha)$ as it can also range on the same scale (see above).

$I_k(\theta)$ aims at measuring the extent of dissimilarity that anonymization introduced for j and j', with $\theta$ capturing different emphasis on relative permutation distances; the greater $\theta$, the stronger the focus on large distances[9]. In a similar fashion to disclosure risk, aversion to information loss accounts for the fact that different agents involved in data dissemination can all have different perceptions of information loss. Typically, this aversion is likely to be stronger for data users than for data releasers. The parameter $\theta$ formalizes such diversity in tastes. As for $D_j(\alpha)$, it can also be used to rank unambiguously couples of anonymization methods (or the same anonymization method with two different parametrizations) that perform better than others, by introducing the concept of dominance in information:

**Definition 4**: *For two attributes j and j', two anonymization methods A and B are said to dominate (i.e. perform better than) two other methods C and D for the preservation of information if it holds that $I_k(\theta)' \leq I_k(\theta)\ \forall\ \theta \geq 1$, where $I_k(\theta)'$ (resp. $I_k(\theta)$) are the measures of Proposition 5 computed on A and B (resp. C and D)).*

Information dominance characterizes anonymization methods that, when applied on two attributes, will consistently ensure lower levels of relative permutation distances (and thus a greater preservation of information) from the mean to the top of their distribution. In practice, that means that whatever the aversion to information loss any agents being involved in data dissemination can have, a dominant couple of methods compared to others will ensure unanimity regarding its performance in term of information preservation.

---
[9] The limit case $I_k(+\infty)$ is strictly equal to the largest relative permutation distance in $\Delta(r_k)$.

Beyond establishing which couple of methods does best in preserving information, $I_k(\theta)$ and information dominance can also be used to tune the extent of information to be preserved. Under different scenario of aversion to information loss, two anonymization methods can be evaluated ex-post in term of information preservation through $I_k(\theta)$ and then re-run to obtain the desired information loss. However, as any anonymization methods can be viewed as functionally equivalent to permutation, nothing precludes the data administrators to select a set of permutation matrices that will target a level of information preservation ex-ante, then applying it to the dataset. The permutation paradigm simplifies the implementation of multivariate scenario and the quantification of information loss in comparison to the current available techniques.

## 5. Extensions
### 5.1 Remarks on local anonymization methods in the permutation paradigm

Multivariate scenario in the permutation paradigm can be viewed as permutation of block of attributes. Similarly, local anonymization methods, which operate locally by adding noise to a subset of records, can be viewed as permutation of blocks of *records*. As an illustration, let's assume that in Table 1 only the second and third records have been masked (Table 4). Clearly, the effects of local anonymization can also be viewed as the result of a set of permutation matrices[10] which in turn can be analyzed using the measures developed above. One can in particular confront local methods against global methods to assess their relative performances in term of disclosure risk and information loss.

**Table 4. Local anonymisation in the permutation paradigm**

| Original dataset X | | | Masked dataset Y | | |
|---|---|---|---|---|---|
| $X_1$ | $X_2$ | $X_3$ | $Y_1$ | $Y_2$ | $Y_3$ |
| 13 | 135 | 3707 | 13 | 135 | 3707 |
| 20 | 52 | 826 | 20 | 57 | 822 |
| 2 | 123 | -1317 | -1 | 122 | 248 |
| 15 | 165 | 2419 | 15 | 165 | 2419 |
| 29 | 160 | -1008 | 29 | 160 | -1008 |

| Rank of the original attribute | | | Rank of the masked attribute | | |
|---|---|---|---|---|---|
| $X_{1R}$ | $X_{2R}$ | $X_{3R}$ | $Y_{1R}$ | $Y_{2R}$ | $Y_{3R}$ |
| 4 | 3 | 1 | 4 | 3 | 1 |
| 2 | 5 | 3 | 2 | 5 | 3 |
| 5 | 4 | 5 | 5 | 4 | 4 |
| 3 | 1 | 2 | 3 | 1 | 2 |
| 1 | 2 | 4 | 1 | 2 | 5 |

| Reverse mapped dataset Z | | | Noise E | | |
|---|---|---|---|---|---|
| $Z_1$ | $Z_2$ | $Z_3$ | $E_1$ | $E_2$ | $E_3$ |
| 13 | 135 | 3707 | 0 | 0 | 0 |
| 20 | 52 | 826 | 0 | 5 | -4 |
| 2 | 123 | -1008 | -3 | 0 | 1256 |
| 15 | 165 | 2419 | 0 | 0 | 0 |
| 29 | 160 | -1317 | 0 | 0 | 309 |

Moreover, the measures introduced in this paper, which all relies on the use of power means, are in fact particularly suitable, as power means have the property of *sub-block consistency* ([7]). In the context of data anonymization, that means that if according to the measures developed above, information loss increases and disclosure risk decreases over some sub-groups of m<n records, while remaining constant on the others n-m records, then the measures of disclosure risk and information loss computed on the n records will also respectively strictly increase and decrease. This coherency makes a compelling argument for the use of the measures developed above.

---

[10] For the first and second attributes it is the identity matrix and for the third it is $\begin{pmatrix} 1 & 0 & 0 & 0 & 0 \\ 0 & 1 & 0 & 0 & 0 \\ 0 & 0 & 0 & 0 & 1 \\ 0 & 0 & 0 & 1 & 0 \\ 0 & 0 & 1 & 0 & 0 \end{pmatrix}$.

### *5.2 Measures of disclosure risk and information loss at the data set level*

The class of disclosure risk measures introduced in Section 3 operates by attributes taken in isolation. While this is a standard approach, one may also be interested in having a quantification of the overall disclosure risk for a data set of p attributes. This kind of measure is in a way complementary to an assessment of disclosure risk attribute by attribute: while the latter is necessary to have a detailed view on the level of protection applied, which is likely to vary according to each attribute's specificity and sensitivity, having a global view of the anonymized data set can be useful, not least for communication purposes. Considering as a starting point the measure $D_j(\alpha)$, which as outlined above bears close similarity with a L(p)-norm (i.e. a vector norm), for a data set with p attributes a possible overall measure can be constructed from a L(p,q)-norm (i.e. a matrix norm, see [6]):

**Proposition 6**: *For a data set $Y_{(n,p)}$, an overall quantitative measure of disclosure risk in the permutation paradigm is given by:*

$$D(\alpha, \beta) = \left[\frac{1}{p}\sum_{j=1}^{p} D_j(\alpha)^\beta\right]^{\frac{1}{\beta}} \text{ for } \alpha \leq 1, \beta \leq 1 \text{ and } \beta \neq 0$$

$$\text{and } D(\alpha, \beta) = \prod_{j=1}^{p} D_j(\alpha)^{1/p} \text{ for } \alpha \leq 1 \text{ and } \beta = 0$$

$D(\alpha, \beta)$ operates in two stages: it first measures disclosure risk for each attributes with $D_j(\alpha)$, then summarizes these p measures into a single one. Equivalently, it first aggregates the columns of the matrix formed by the collection of the p vectors of rank displacements $r_j$ and then aggregates the p measures. $N(\alpha, \beta)$ is based on the expression of a L(p,q)-norm but doesn't define one strictly due to the $\sqrt[\alpha]{n}$ and $\sqrt[\beta]{p}$ factors and also the range of variation of $(\alpha; \beta)$: following the same reasoning than for $\alpha$ in $D_j(\alpha)$, $\beta$ is set to range from one and below. This constraint is attached to the interpretation that can be given to the parameter $\beta$ in the context of data anonymization. $\beta = 1$ is the benchmark case where all attributes in the data set are weighted equally: from a disclosure risk perspective, all attributes matter the same way. But when $\beta$ decreases, more weight is given to the lowest protected attributes in the dataset; in the limit case with $\beta \to -\infty$, the overall disclosure risk of the data set is assessed through the perspective of the least protected attribute (i.e. the one having the lowest $D_j(\alpha)$ value). As for $\alpha$ in $D_j(\alpha)$, $\beta$ in $D(\alpha, \beta)$ substantiates the variety of preferences in disclosure risk that users or releasers can have, but here this variety is expressed across attributes in the context of an overall diagnosis of disclosure risk for a data set.

Along the same line, it can be constructed an overall measure of information loss for a data set. Assuming that if in $Y_{(n,p)}$ its p attributes are to be masked, there are $j(j-1)/2$ potential sources of information loss (i.e. k distinct couples of attributes). Aggregating all these sources can be done by taking the norm of the matrix formed by the collection of the $j(j-1)/2$ relative permutation distances vectors $\Delta(r_k)$, which gives:

**Proposition 7**: *For a data set $Y_{(n,p)}$ with p attributes to be protected against disclosure risk, an overall quantitative measure of information loss in the permutation paradigm is given by:*

$$I(\theta, \pi) = \left[\frac{1}{j(j-1)/2} \sum_{k=1}^{j(j-1)/2} I_k(\theta)^\pi\right]^{\frac{1}{\pi}} \text{ for } \theta \geq 1 \text{ and } \pi \geq 1$$

$I(\theta, \pi)$ operates also in two stages: it first measures information loss for every possible distinct couples of attributes, then summarizes these $j(j-1)/2$ measures into a single one. Equivalently, it first aggregates the columns of the matrix formed by the $j(j-1)/2$ vectors of *relative* rank displacement $\Delta(r_k)$ and then aggregates the collection of $j(j-1)/2$ measures. $I(\theta,\pi)$ is also based on the expression of a L(p,q)-norm and in fact does define one up to the $\sqrt[\theta]{n}$ and $\sqrt[\pi]{j(j-1)/2}$ factors. In particular, the range of variation of $\pi$ is interpretable in term of information loss. $\pi = 1$ is the benchmark case where every attributes in the data set are weighted equally and matter the same way in term of information loss. When $\pi$ increases, more weight will be given to the couple of attributes with the largest information loss; in the limit case with $\pi \to +\infty$, the overall information loss of the data set is assessed through the perspective of the least preserved couple of attribute (i.e. the ones having the highest $I_k(\theta)$ value). As for $\theta$ in $I_k(\theta)$, $\pi$ in $I(\theta,\pi)$ substantiates the variety of preferences in information loss that users or releasers can have, but here such variety is expressed across attributes in the context of an overall diagnosis of information loss for a data set.

### 5.3 Remarks on intruder models and the Kerckhoff's principle

In the context of the permutation paradigm, it is considered in [3] a scenario of known-plaintext attack in anonymization, which defines an intruder who knows both the original data set and its entire corresponding anonymized version. This a rather extreme configuration, unlikely to be mirrored by concrete situations, not least because the intruder has nothing to gain except the disclosure of the method used by the data protector. However, at first glance it is conceptually very insightful, as anonymization that can pass the test of such situation will in fact be able to pass any test.

Under this scenario, the attacker can eliminate the small noise matrix of *Proposition 1 (*[3]). The random seed for the noise is thus irrelevant, which leaves to uncover the random seed for permutation. As the intruder has maximum knowledge, some permutation matrices can also be retrieved easily. But they can only be random in nature, as for an attribute j with n records, n! possible permutation matrices are possible and only one of them will be the true $P_j$. This result streams from the fact that the data releaser knows, through the ID attached to each records, which particular collection of attribute values in the disseminated, anonymized data, is linked to a specific collection of attribute values in the original data. The attacker has no knowledge of this and the purpose of the attack is to uncover the exact $P_j$'s, as only them will allow to show which anonymized records derive from which original record values.

This leads to two consequences. An immediate, trivial one is that the task of an intruder will be harder on larger dataset. Thus, the permutation paradigm makes explicit the fact that large micro data happen to be better suited for protection than small ones. A second consequence is that SDC can spare to itself to work under the "security through obscurity" principle which is generally embraced by cryptographers. With the permutation paradigm, it is evident that the use of SDC techniques can be conceptually viewed as secure cryptosystems even if everything about the system, except the key, is public knowledge. The key happens to be the collection of $P_j$'s or, equivalently, of rank displacements matrices.

Further exploration of the Kerckhoff's principle in future research are however required under less stringent, and thus more realistic, intruder scenario, and in particular appropriate and realistic background knowledge assumptions about the intruder. We argue that the concepts of aversion to disclosure risk and information loss developed in this paper are a step in the right direction, insofar as they allow conveying information on the way data have been anonymized for different preferences.

### 6. Conclusions and future research

As rightfully recognized by its original authors, the permutation paradigm opens several new directions in data anonymization. This paper has tried to explore some them.

First, by a simple restatement of the permutation paradigm, we have characterized permutation matrices as an encompassing tool in data anonymization. Based on this observation, we have derived two general classes of disclosure risk and information loss measures, which we argued are easy to compute for every methods and data sets and are meaningful. These two classes are based on the aggregative structure of p-norms (albeit they don't always define p-norms strictly), and the degrees of these norms can be harnessed with an interpretation in term of aversion. In the case of disclosure risk, the aversion translates in different emphasis on the lowest permutation distances achieved among records for one attribute. For information loss, the aversion translates in different emphasis on the highest relative permutation distances among records between two attributes. While every data releasers and users alike would like to achieve the unfeasible ideal of data with maximum protection against disclosure risk and minimal information loss, in practice they all have different judgments and utility/risk trade-offs. The measures developed in this paper allow to incorporate such diversities and also importantly to communicate on them. In addition, they allow deriving unanimity of judgments following the concepts of dominance introduced in this paper. The incorporation of these judgments can be viewed as a step toward the application of the Kerckhoff's principle in data anonymization under realistic intruder scenario.

This paper opens new future research lines in the permutation paradigm and data anonymization in general, including:

- Establish an inventory of popular methods under different parametrizations and data contexts, using the class of measures developed in this paper, in particular for benchmarking the values of these measures into existing practices
- Characterize the methods that are dominant in term of disclosure risk and information loss
- In particular, establish in practice if some methods can be both dominant in terms of disclosure risk and information loss, which could provide a strong rational for their uses
- Following the development of co-utility games in data anonymization ([12], [2]), determine if dominant equilibriums can be reached in these models
- Elaborate some graphical tools for the display of disclosure risk and information loss, in particular the representation through dominance curves
- Explore further the possibility of using permutation matrices as direct input to data anonymization